# Markup Language Modeling for Web Document Understanding

Su Liu, Bin Bi, Jan Bakus, Paritosh Kumar Velalam,
Vijay Yella, Vinod Hegde

## Abstract

Web information extraction (WIE) is an important part of many e-commerce systems, supporting tasks like customer analysis and product recommendation. In this work, we look at the problem of building up-to-date product databases by extracting detailed information from shopping review websites. We fine-tuned MarkupLM on product data gathered from review sites of different sizes and then developed a variant we call MarkupLM++, which extends predictions to internal nodes of the DOM tree. Our experiments show that using larger and more diverse training sets improves extraction accuracy overall. We also find that including internal nodes helps with some product attributes, although it leads to a slight drop in overall performance. The final model reached a precision of 0.906, recall of 0.724, and an F1 score of 0.805.

## 1 Introduction

The web contains vast amounts of text and multimedia content, making it the largest platform for communication and the exchange of opinions (Varlamov and Turdakov, 2016). The data from web has several applications such as providing improved search results and construction of databases to serve user queries (Dhillon et al., 2011). Every day, the vast and rapidly increasing amount of information is constantly being produced, shared, and consumed online. To leverage the web data effectively, we must efficiently apply Web Information Extraction (WIE) with minimal human effort (Ferrara et al., 2014).

WIE differs from traditional IE by processing semi-structured online documents (Bing et al., 2013). It leverages on scientific methods coming from various disciplines including tag path clustering, Bayesian, text mining, rule based approach, machine learning, wrappers family algorithm and others (Ferrara et al., 2014; Chang et al., 2006; Bin Mohd Azir and Ahmad, 2017).

WIE presents several challenges. One key challenge is automating WIE with minimal human intervention while maintaining high accuracy (Doan et al., 2009). Additionally, it must handle large volumes of data quickly, especially in business and competitive intelligence contexts, and ensure privacy protection, particularly when dealing with personal data (Abdullah et al., 2023). Furthermore, if using machine learning approaches, it often requires large, manually labeled datasets, which are time-consuming to get and prone to errors (Abdullah et al., 2023; Li et al., 2023; Dhillon et al., 2011). The evolving nature of web data sources also poses a significant challenge, as structural changes can occur unpredictably, necessitating flexible and maintainable systems to adapt to these changes (Hong et al., 2021).

Although there are several challenges, WIE has become a crucial tool in many fields, including e-commerce, where vast amounts of product and user data are continuously generated across numerous online platforms. WIE can help to develop various applications (Bing et al., 2013), such as recommendation and question answering. To build these applications, product attributes need to be extracted from web accurately.

Depending on the structure of HTML, the product item pages can be broadly categorized into detail pages and list pages. Detail pages provide detailed information about a single item; while list pages comprise a list of items with abridged detail, organized under a single theme (San et al., 2023). Some product extraction datasets are about detail pages, such as SWDE (Hao et al., 2011), the Klarna Product Page Dataset (Hotti et al., 2024), and WDC (Petrovski et al., 2017). Some are about list pages, including BSCM (Dhillon et al., 2011), PLAtE (San et al., 2023), and datasets created by Furche (Furche et al., 2012). Our research focuses on extracting web information from both detail pages and list pages. By doing so, we eliminate the need to distinguish between these two types of pages prior to extraction, allowing us to gather information from a wider range of web domains.

In addition to handling different page types, our WIE method is designed to extract highly detailed and frequently updated product information. While there are existing product information extraction databases, such as Open Products Data (Arribas-Bel et al., 2021), Products-10k (Bai et al., 2020), GreenDB (Jäger et al., 2022), these are tailored for specific purposes. Open Products Data is community-driven, Products-10k is optimized for image-based tasks, and GreenDB specializes in sustainability-related product data. Furthermore, many product datasets suffer from infrequent updates, leading to outdated information (Flick et al., 2023). In contrast, our method ensures that we access the most current data with minimal human effort.

In conclusion, there is a clear demand for a more robust and efficient method capable of extracting the most up-to-date, detailed, and comprehensive product information from a wide range of shopping review websites.

## 2 Related Work

At the beginning, a set of wrappers were created to extract unstructured information into structured one (Bin Mohd Azir and Ahmad, 2017; Gulhane et al., 2011). Then, some ML based methods, such as Naïve Bayes (Freitag, 2000), support vector machine (Bhuyan and Chakraborty, 2024), have also been applied to WIE. The modern methods of WIE are based on deep learning, aiming to extract information with deep semantic understanding (Li et al., 2023). Based on backbone network architecture, these modern approaches can be divided into graph neural network-based methods (Kipf and Welling, 2017; Kumar et al., 2022; Hwang et al., 2021) and transformer-based methods (Deng et al., 2022; Li et al., 2022; Wang et al., 2022).

Nowadays, there are some transformer-based models utilize neural-based approaches to construct a representation for each HTML node for the extraction task. RoBERTa (Liu et al., 2019) was pre-trained on natural language texts from BookCorpus (Zhu et al., 2015) and Wikipedia using masked language modeling, which serves as the foundation for several such models.

DOM-LM (Deng et al., 2022), built on RoBERTa, generates contextualized representations for HTML documents by pre-training on the SWDE dataset (Hao et al., 2011), using masked language modeling on text tokens and DOM tree nodes.

MarkupLM (Li et al., 2022), another RoBERTa-based model, processes a node represented by both an XPath embedding and its text. The XPath embedding is created by embedding each tag and subscript separately, concatenating them, and passing them through a feed-forward neural network. MarkupLM was pre-trained on masked markup language modeling, node relation prediction, and title page matching tasks using 24 million pages from Common Crawl.

Then, WebFormer (Wang et al., 2022) designs several DOM structure based attention mechanisms, effectively leveraging web layout for superior performance on SWDE and Common Crawl benchmarks. WIERT (Li et al., 2023) was designed to overcome several disadvantages the former models have, such as the input length limit of 512 tokens.

In this study, we chose the MarkupLM model because it is designed to make predictions solely on the leaf nodes of a DOM tree, using a sequential labeling approach. An example is illustrated in Figure 1 below.

As shown to the left, given the HTML page and the corresponding DOM tree, the input sequence to MarkupLM would only consists of the text in the leaf nodes (highlighted in red boxes). As a result, the MarkupLM can only make predictions on these leaf nodes.

However, predictions on internal nodes (HTML tags such as <div> and <li> etc.) are needed for productization, since we want to know something like which node a subtree is rooted at corresponds to a certain product. There were a couple of heuristic methods trying to push up the predictions from leaf nodes to internal nodes, but they are not reliable enough.

Apart from fine-tune the MarkupLM using product data, we also worked on a new principal way for MarkupLM to make predictions directly on internal nodes. We name the new method MarkupLM++. The idea is inserting the internal nodes (HTML tags) to the input text to MarkupLM as new tokens. This allows the model to make predictions on these new tokens. For example, here is the new sequence input to MarkupLM++ for the example webpage illustrated, where HTML tags like <head> and <title> are added:
[CLS] <html> <head> <title> Galaxy S20 [SEP] <body> <div> <li> <div> <span> Display 6.5 inch ......

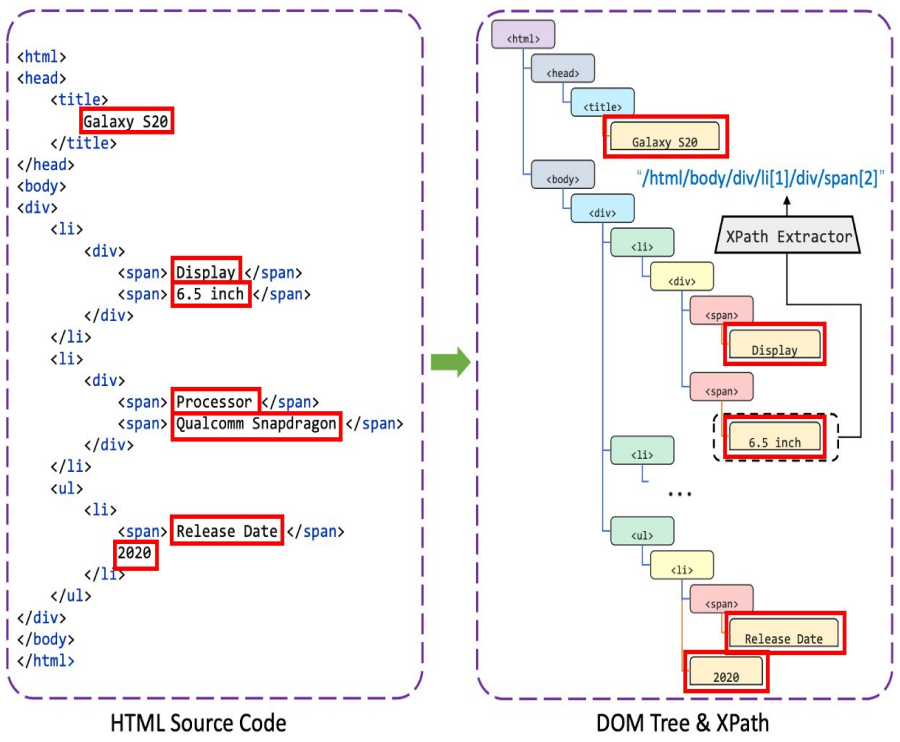

Figure 1: An example of DOM tree and XPath with the source HTML code (Li et al., 2021)

In original MarkupLM, given the difference in levels between predictions and annotations (predictions at the token level vs. annotations at the node level), original MarkupLM had to aggregate token-level predictions on the text of a node to derive a prediction for this node in some way (max or mean), in order to match with the annotation. In contrast, the MarkupLM++ is trained by minimizing the multi-label cross-entropy loss only on nodes (tags) instead of tokens. Therefore, the new model does not suffer the level mismatch problem, as it makes predictions directly on nodes which are consistent with annotations.

## 3 Methodology

The HTML data used for fine-tuning and testing was gathered from various online shopping review websites. A preceptor tool, integrated within the crawler pipeline, labeled the product information using heuristic methods. The preceptor tool can only extract product data from a limited number of web domains, as it relies on domain-specific heuristics.

We labeled the product information using 17 attributes: article paragraph, headline, product brand, product cons, product cons label, product container, product hook, product link, product name, product ordinal, product price, product pros, product pros label, product review, product seller, product bottom line, and product bottom line label. The article paragraph refers to the introduction section of a web article, typically a brief overview of the product placed under the headline or sub headline. The product container represents the frame that holds the product information. Product ordinal indicates the product's position when multiple products are listed on the same webpage. The product bottom line and its label refer to the final piece of information on the product webpage. Not all categories are present on every webpage. The three most important categories for us are product name, product link, and product review.

We used training sets of different sizes to evaluate whether the model's performance improves. Three datasets were used for fine-tuning MarkupLM: one containing 17 domains, one with 48 domains, and one with 84 domains. The models were named according to the number of domains used for fine-tuning. For example, MarkupLM p-17 indicates

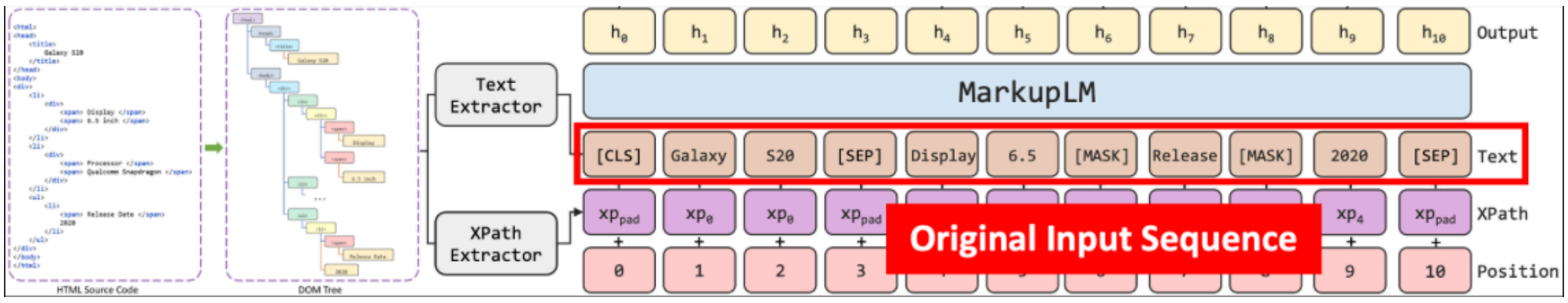

Figure 2: The architecture of MarkupLM, where the pre-training tasks are also included (Li et al., 2021)

that the model is fine-tuned and testing using product datasets from 17 domains.

### 3.1 MarkupLM p-17

The first dataset consists of 17 product domains. 4 domains were reserved for the testing set, which included 12,821 web pages from sites such as autoguide.com, bikeperfect.com, livescience.com, and tomsguide.com. The remaining 13 domains, containing 37,563 web pages, were used to fine-tune the model.

### 3.2 MarkupLM p-48

The dataset included 596,302 samples from 48 product review domains. To evaluate the model's ability to generalize to new, unseen domains, we reserved 5 domains as testing data, which contained 29,860 samples. These 5 test domains were not included in the fine-tunning data, offering a robust assessment of the DU model's performance on new domains with zero-shot annotation in the future.

### 3.3 MarkupLM++

The MarkupLM++ model was fine-tuned on the same datasets as its predecessor, MarkupLM p-48. We reserved 5 domains as testing data. However, unlike the previous version, where all internal nodes in HTML trees were pruned, the modified algorithm retained these nodes.

After we collected the predictions from these models, we applied human annotation to double check whether the predictions were valid.

## 4 Results

In our case, we place more emphasis on the micro-average of the metrics, as we prioritize specific classes, such as product name, product link, and product review, which also have a larger number of samples.

Table 1 below presents the micro-averages of the metrics for the three models. For precision, recall, and F1 score, MarkupLM p-48 consistently outperforms the others. MarkupLM p-17 and MarkupLM++ show similar metrics, both falling below the performance of MarkupLM p-48.

Additionally, for all three models, precision consistently exceeds recall, indicating that while the models accurately predict product information, some details are still being missed.

Table 2 presents the performance of the three models on the most frequently used classes. For product name, MarkupLM p-48 provides the best predictions. However, for product link, it shows low recall, indicating that many product links were not identified. Upon investigation, we discovered that the model predominantly predicts on leaf nodes, but in some cases, these leaf nodes are structured as '…/a/span' rather than the expected '…/a'. As a result, the product links are in internal nodes, which are pruned by MarkupLM p-48. This explains why MarkupLM++, which also predicts on intermediate nodes, has a higher recall for product links. For MarkupLM p-17, the recall for product links is higher than that of MarkupLM p-48, as the training and testing samples are more consistent, with fewer instances of the '.../a/span' structure in the testing sets. For product review, MarkupLM++ outperforms the other models, while MarkupLM p-17 shows significantly lower results.

| **Model** | **P** | **R** | **F1** |
|---|---|---|---|
| MarkupLM p-17 | 0.802 | 0.697 | 0.746 |
| MarkupLM p-48 | 0.906 | 0.724 | 0.805 |
| MarkupLM++ | 0.819 | 0.663 | 0.733 |

Table 1: Precision, recall and F1 micro-average of all models.

| **Model** | **Attribute** | **P** | **R** | **F1** |
|---|---|---|---|---|
| MarkupLM p-17 | Product name | 0.711 | 0.632 | 0.669 |
| | Product link | 0.494 | 0.596 | 0.540 |
| | Product review | 0.773 | 0.820 | 0.796 |
| MarkupLM p-48 | Product name | 0.738 | 0.676 | 0.706 |
| | Product link | 0.933 | 0.339 | 0.497 |
| | Product review | 0.880 | 0.938 | 0.908 |
| MarkupLM ++ | Product name | 0.615 | 0.512 | 0.559 |
| | Product link | 0.734 | 0.632 | 0.680 |
| | Product review | 0.928 | 0.917 | 0.922 |

Table 2: Precision, recall and F1 of all models on the key attributes.

### 4.1 MarkupLM p-17

The model performs well in extracting key fields such as headline and product_review, with precision and recall for headline being exceptionally high (0.982 and 0.98.4, respectively). The metrics of product_brand and product_seller are zero. Since all the samples are true negative, the domains used for testing do not contain such information. Product_review shows good recall (0.82) but slightly lower precision (0.773), suggesting the model is extracting some irrelevant text as reviews.

While the model performs reasonably for product_link and product_container, it generally has higher recall than precision, meaning it identifies more relevant information but also includes false positives. Classes such as product_bottom_line shows very low performance, indicating difficulties in identifying these fields.

The Micro-Average (0.746 F1) is higher than the Macro-Average (0.406 F1), suggesting the model performs better on larger, more frequent classes. The Macro-Average is impacted by poor performance in less frequent categories. In conclusion, the model excels at common fields but needs improvement in less frequently seen categories.

| Attribute | P | R | F1 |
|---|---|---|---|
| article_paragraph | 0.882 | 0.632 | 0.736 |
| headline | 0.982 | 0.984 | 0.983 |
| product_brand | 0.000 | 0.000 | 0.000 |
| product_cons | 0.491 | 0.552 | 0.520 |
| product_cons_label | 0.204 | 0.177 | 0.190 |
| product_container | 0.887 | 0.671 | 0.764 |
| product_hook | 0.108 | 0.203 | 0.141 |
| product_link | 0.494 | 0.596 | 0.540 |
| product_name | 0.711 | 0.632 | 0.669 |
| product_ordinal | 0.213 | 0.928 | 0.347 |
| product_price | 0.451 | 0.819 | 0.582 |
| product_pros | 0.489 | 0.448 | 0.468 |
| product_pros_label | 0.185 | 0.147 | 0.164 |
| product_review | 0.773 | 0.820 | 0.796 |
| product_seller | 0.000 | 0.000 | 0.000 |
| product_bottom_line | 0.002 | 0.007 | 0.003 |
| product_bottom_line_label | 0.000 | 0.000 | 0.000 |
| Macro-Average | 0.404 | 0.448 | 0.406 |
| Micro-Average | 0.802 | 0.697 | 0.746 |

Table 3: Precision, recall, and F1 of MarkupLM p-17 by attributes.

### 4.2 MarkupLM p-48

For the models fine-tuned using the 48 domain datasets, the predictions of headline and product_container perform particularly well, and product_container has an improvement comparing to MarkupLM p-17. Product_review shows an obvious improvement as well.

In the larger dataset, there is information about product brand. However, it has high precision (0.939) but low recall (0.589), resulting in F1 score of 0.724. Product_link also has high precision (0.933) but a low recall (0.339), meaning the model identifies product links accurately but misses many. Product_name shows moderate performance and only a slight increase comparing to MarkupLM p-17.

Some attributes, such as product_bottom_line_label, show a performance of zero. Since all the samples are true negatives, this indicates that there is no product information of that type in the domains. Product_cons and product_pros also show challenges, with high precision but low recall, leading to low F1 scores, meaning identifying pros and cons of products is challenging to the model.

The Micro-Average F1 score (0.805) is significantly higher than the Macro-Average

| Attribute | P | R | F1 |
|---|---|---|---|
| article_paragraph | 0.770 | 0.896 | 0.828 |
| headline | 0.998 | 0.838 | 0.911 |
| product_brand | 0.939 | 0.589 | 0.724 |
| product_cons | 0.833 | 0.187 | 0.306 |
| product_cons_label | 0.835 | 0.411 | 0.551 |
| product_container | 0.965 | 0.806 | 0.878 |
| product_hook | 0.972 | 0.602 | 0.743 |
| product_link | 0.933 | 0.339 | 0.497 |
| product_name | 0.738 | 0.676 | 0.706 |
| product_ordinal | 0.232 | 0.064 | 0.100 |
| product_price | 0.750 | 0.822 | 0.785 |
| product_pros | 0.910 | 0.187 | 0.310 |
| product_pros_label | 0.880 | 0.437 | 0.584 |
| product_review | 0.880 | 0.938 | 0.908 |
| product_seller | 0.004 | 0.695 | 0.009 |
| product_bottom_line | 0.000 | 0.001 | 0.000 |
| product_bottom_line_label | 0.000 | 0.000 | 0.000 |
| Macro-Average | 0.685 | 0.500 | 0.520 |
| Micro-Average | 0.906 | 0.724 | 0.805 |

Table 4: Precision, recall, and F1 of MarkupLM p-48 by attributes.

(0.52), suggesting that the model performs better on more frequent attributes. Compared to MarkupLM p-17, both the macro-average and micro-average show substantial improvement, indicating that fine-tuning with a larger product data size enhances the model's performance.

### 4.3 MarkupLM++

Table 5 highlights the performance of MarkupLM++. It works well at extracting common product information, performing best on product review and product link, though there is still room for improvement in extracting product links. However, it performs the worst when extracting product name.

Among the 17 attributes, MarkupLM++ performed the best on 8 attributes, while MarkupLM p-48 excelled on 7 attributes. It performed particularly well on product pros and cons, as well as their labels, outperforming the other two models. However, it performed poorly on product brand. Like the other models, it predicted more effectively on the more frequent attributes, although its precision, recall, and F1 scores were all lower than those of MarkupLM p-48.

There are certain attributes that all three models struggled to predict, such as product bottom line, product ordinal, and product seller. Upon further investigation, we found that most samples for product seller and product bottom line were false positives and true negatives, indicating that these attributes were less represented in the fine-tuning data. In some cases, product brands were also listed as the sellers. For product ordinal, which only appears on list pages, models fine-tuned with larger datasets did not outperform MarkupLM p-17.

| **Attribute** | **P** | **R** | **F1** |
|---|---|---|---|
| article_paragraph | 0.532 | 0.285 | 0.372 |
| headline | 0.975 | 0.973 | <u>0.974</u> |
| product_brand | 0.000 | 0.000 | 0.000 |
| product_cons | 0.654 | 0.730 | <u>0.690</u> |
| product_cons_label | 0.776 | 0.960 | <u>0.858</u> |
| product_container | 0.854 | 0.603 | 0.707 |
| product_hook | 0.003 | 0.001 | 0.002 |
| product_link | 0.734 | 0.632 | <u>0.680</u> |
| product_name | 0.615 | 0.512 | 0.559 |
| product_ordinal | 0.395 | 0.178 | 0.246 |
| product_price | 0.444 | 0.263 | 0.330 |
| product_pros | 0.570 | 0.826 | <u>0.675</u> |
| product_pros_label | 0.639 | 0.868 | <u>0.736</u> |
| product_review | 0.928 | 0.917 | <u>0.922</u> |
| product_seller | 0.000 | 0.000 | 0.000 |
| product_bottom_line | 0.099 | 0.530 | <u>0.167</u> |
| product_bottom_line_label | 0.000 | 0.000 | 0.000 |
| Macro-Average | 0.483 | 0.487 | 0.466 |
| Micro-Average | 0.819 | 0.663 | 0.733 |

Table 5: Precision, recall, and F1 of MarkupLM++ by attributes.

## 5 Discussion

### 5.1 Adding internal nodes

In the results above, the MarkupLM++ model has a precision of 0.82, which is lower than that of the MarkupLM p-48 model. One possible reason for this is that by retaining the internal nodes, the number of nodes to be predicted increases exponentially. As a result, with a similar size of fine-tuning data, the data becomes sparse, which reduces accuracy.

Another factor could be the pre-trained MarkupLM model itself. Without pre-training on internal nodes, the model does not perform well, even after fine-tuning.

However, for certain classes, especially those with content in internal nodes—such as product links located in '…/a/span' rather than '…/a' or product reviews containing hyperlinks—MarkupLM++ performs better.

Although MarkupLM++ did not outperform MarkupLM p-48 on average metrics, it excelled in predicting the most attributes (8), indicating that adding internal nodes back helps predict content within those nodes. However, this addition also introduces challenges that can reduce overall model performance.

### 5.2 Post-processing

In addition to modifying the structure of the model, we applied engineering methods to address several issues we identified. One such improvement was the addition of a post-processing pipeline.

First, we noticed a discrepancy between the overall precision and the partial precision for product_link, as the model sometimes extracted elements with XPath endings like '/a/span' instead of '/a'. To address this, we removed the '/span' in the post-processing step.

For product_review, the model occasionally extracted not only the reviews but also text containing hyperlinks or formatted content, which was mistakenly identified as part of the review. We resolved this by eliminating those nested extractions, ensuring that only the actual product review text was captured.

Lastly, post-processing also involved grouping predictions by product in the list pages. For example, when a product webpage contained product names, links, and reviews for three different products, the predictions were organized into three separate groups, each corresponding to a specific product.

## 6 Conclusion

Fine-tuning the pre-trained MarkupLM model with HTML data from shopping review webpages can improve its performance, and using a larger dataset with more domains and samples further boosts the model's accuracy. We also experimented with adding the internal nodes back in the pre-trained MarkupLM to enhance predictions on those nodes. While this approach resulted in a decrease in average performance, it improved predictions for certain product attributes, especially those involving internal nodes. By combining the strengths of both models, we can create a detailed and frequently updated product dataset.

# A Appendix. Additional Figures.

Below is the loss when we trained the MarkupLM p-48. As shown in the plot, the loss keeps decreasing during the fine-tunning process, suggesting that the model can learn the extraction patters.

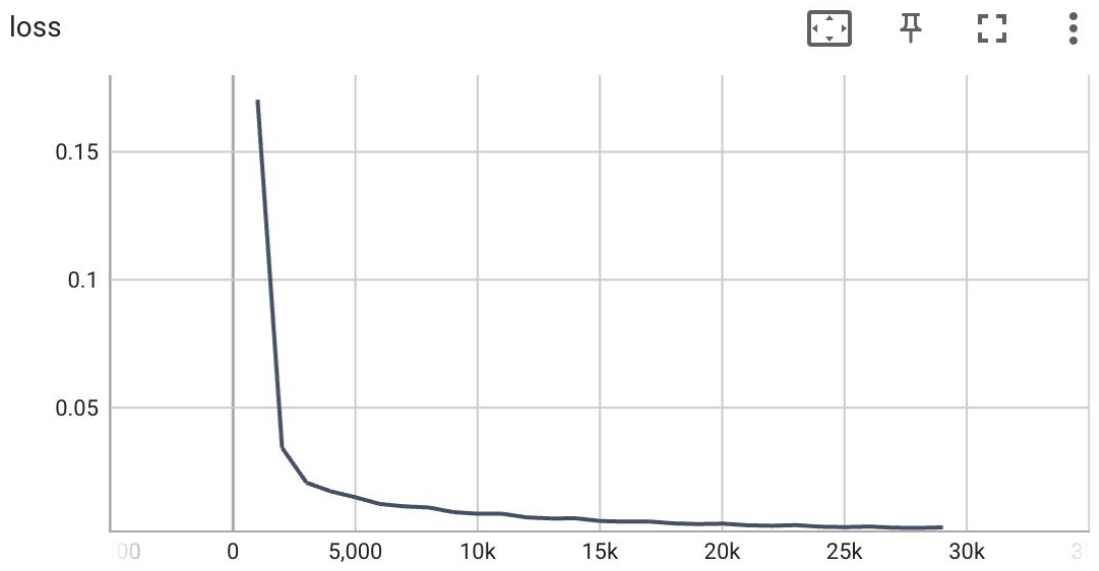

Figure 3: the loss during the fine-tunning process of MarkupLM p-48.

There are also additional figures of good cases and bad cases predicted by MarkupLM p-48.

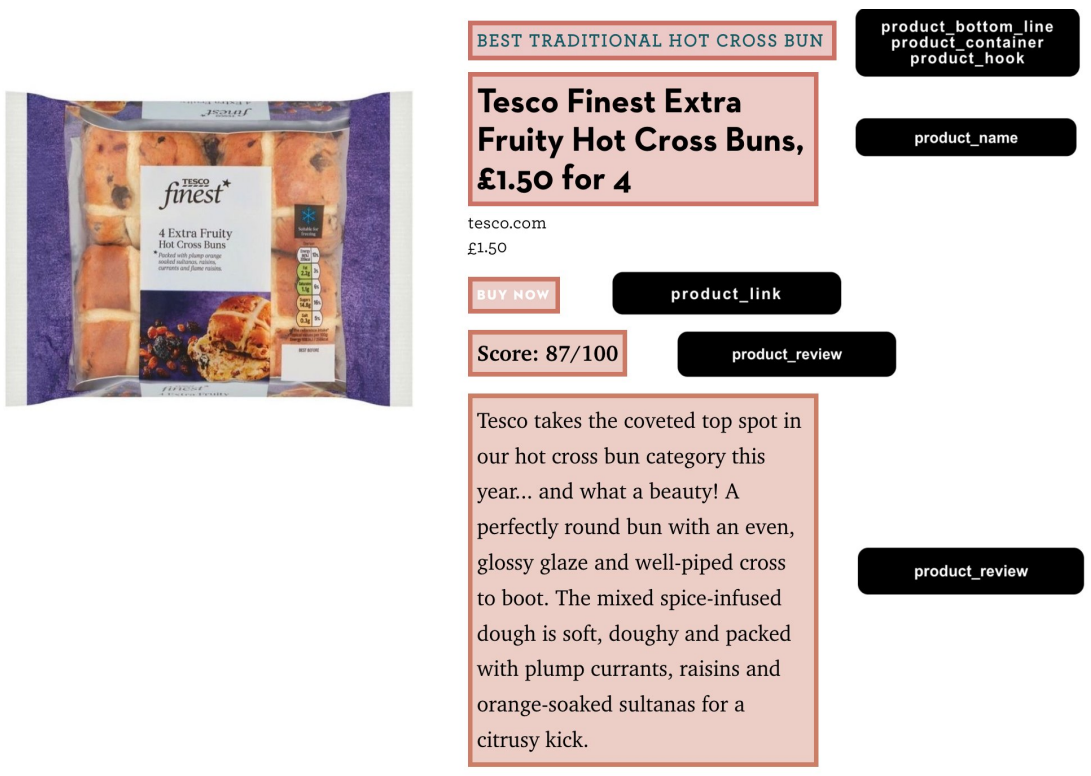

Figure 4: a good case. product name, product link, product review, and product container are extracted correctly.

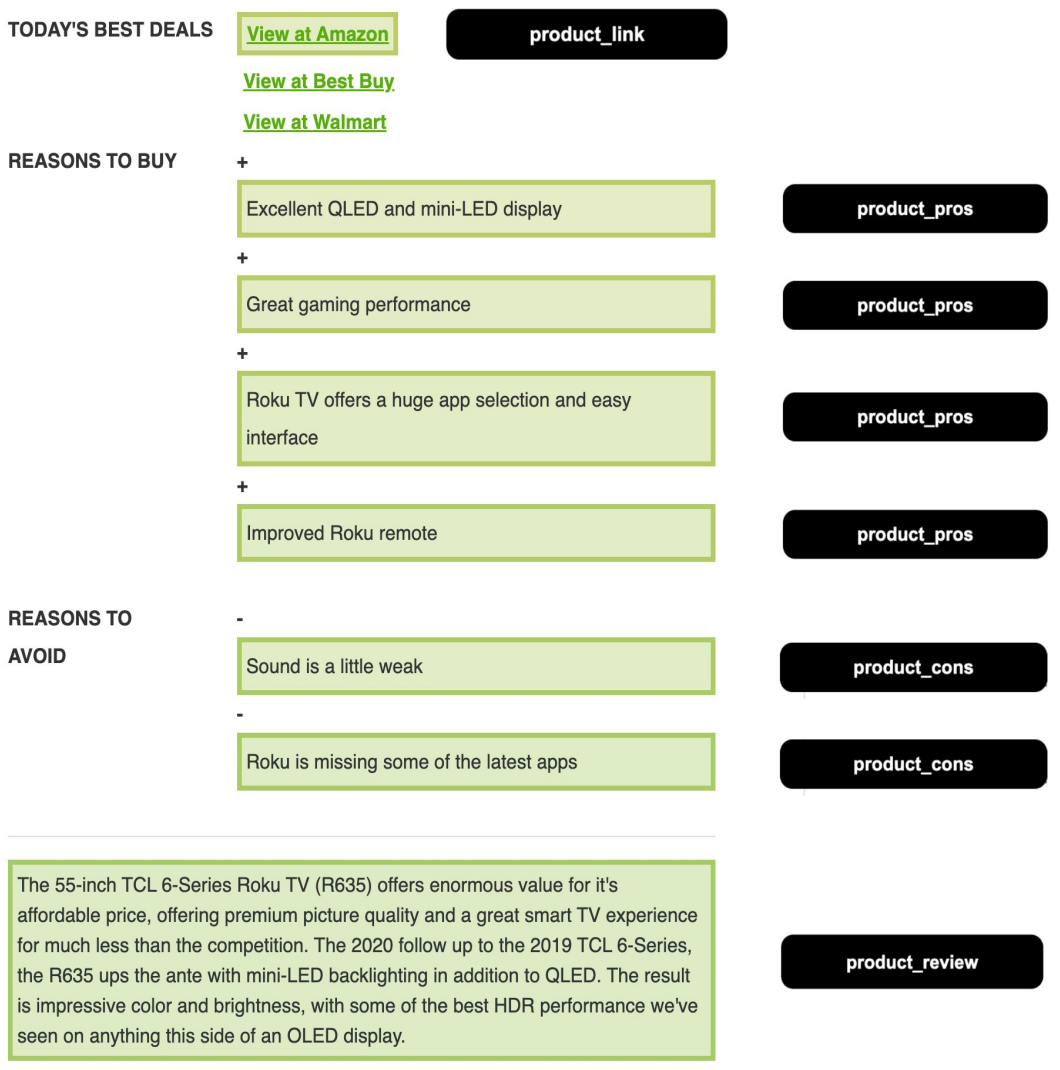

Figure 5: a good case. product pros and product cons are extracted correctly based on the contextual structure (reasons to buy '+', reasons to avoid '-) and textual content. Product link and product review are recognized correctly as well.

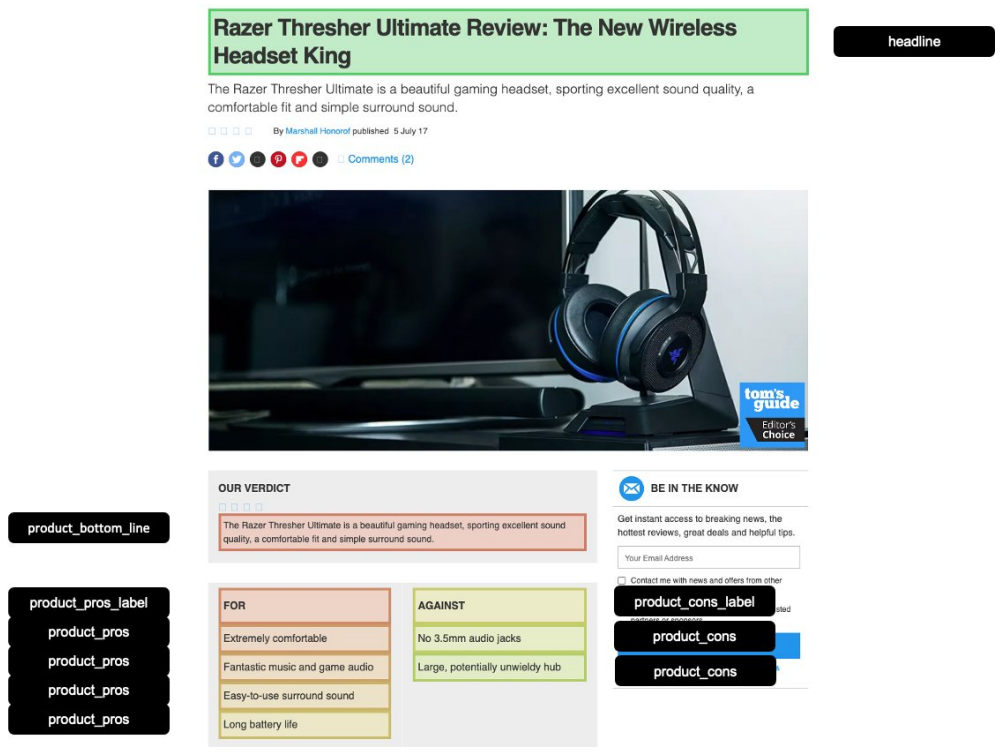

Figure 6: a good case. headline, product bottom line, product pros label, product pros, product cons label, product cons are extracted correctly. The model successfully recognizes "FOR" and "AGAINST" as key indications to extract pros and cons of a product.

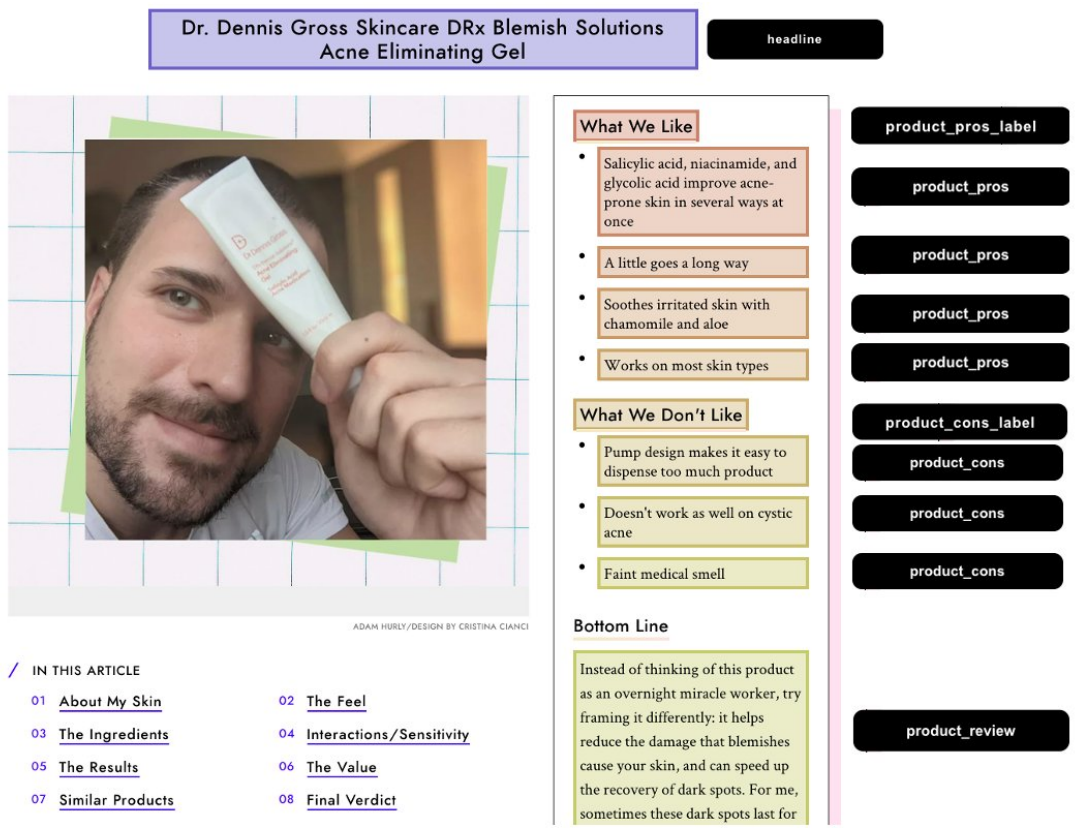

Figure 7: a good case. headline, product pros label, product pros, product cons label, product cons, product review are extracted correctly. The model successfully learns to understand that “What We Like” and “What We Don’t Like” indicate pros and cons of a product, respectively.

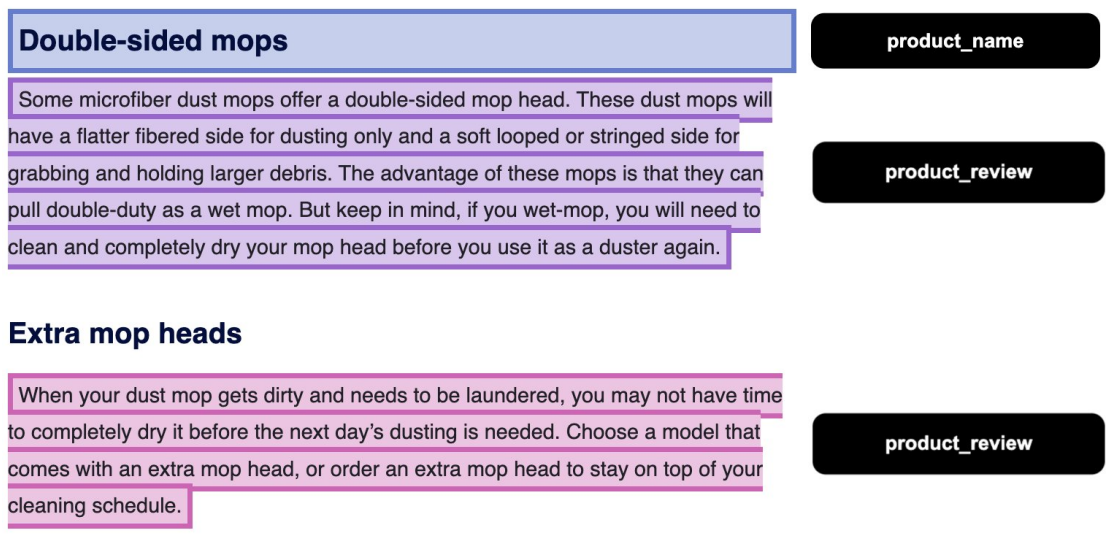

Figure 8: a bad case. While product reviews are recognized correctly, product name is mis-identified, as the “Double-sided mops” refers to a special feature of mops here, instead of an actual

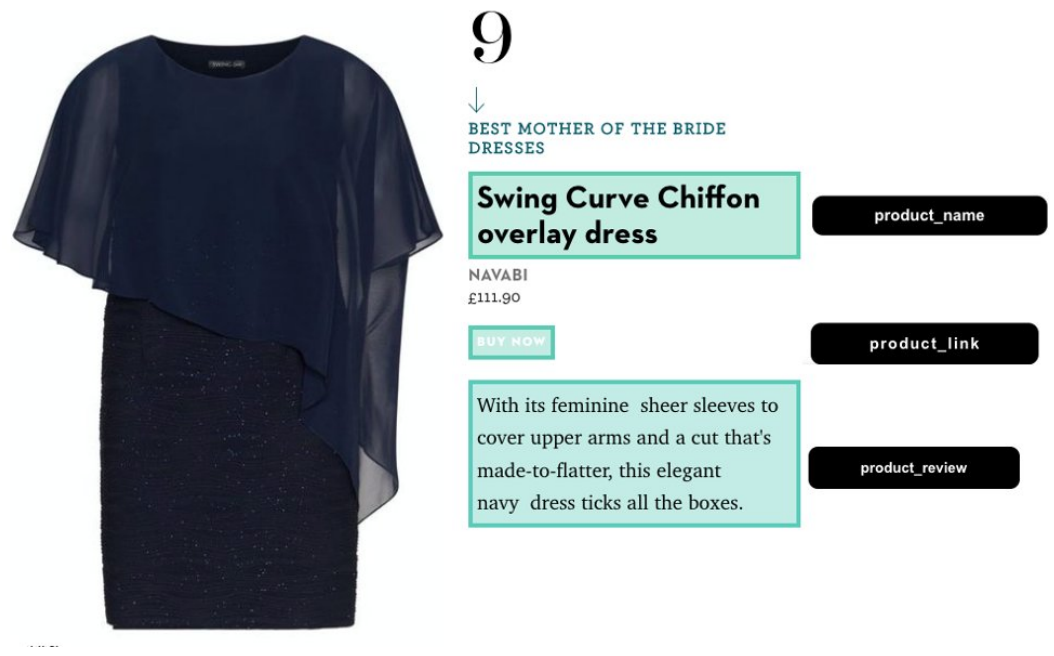

Figure 10: a bad case. The product brand ‘NAVABI’ and product price ‘£111.90’ are not recognized successfully, which lowers the recall.

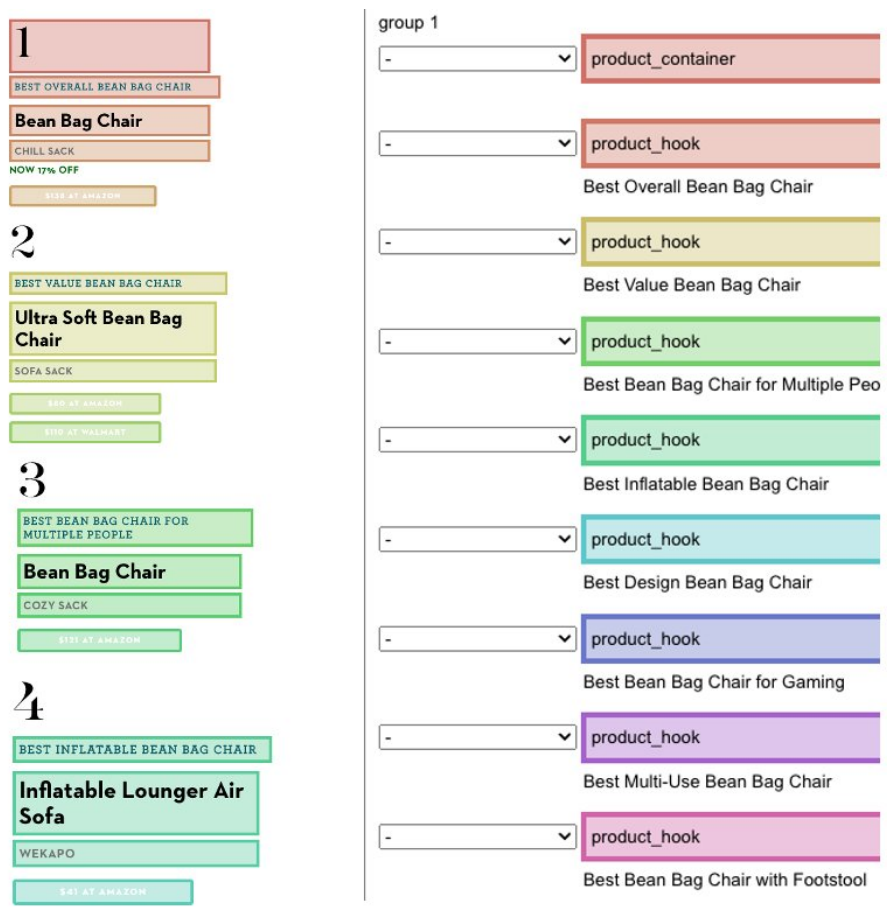

Figure 11: a bad case. The attributes of all the products have been correctly identified, but they are not grouped by product. This issue will be resolved by updated MarkupLM++ described below.

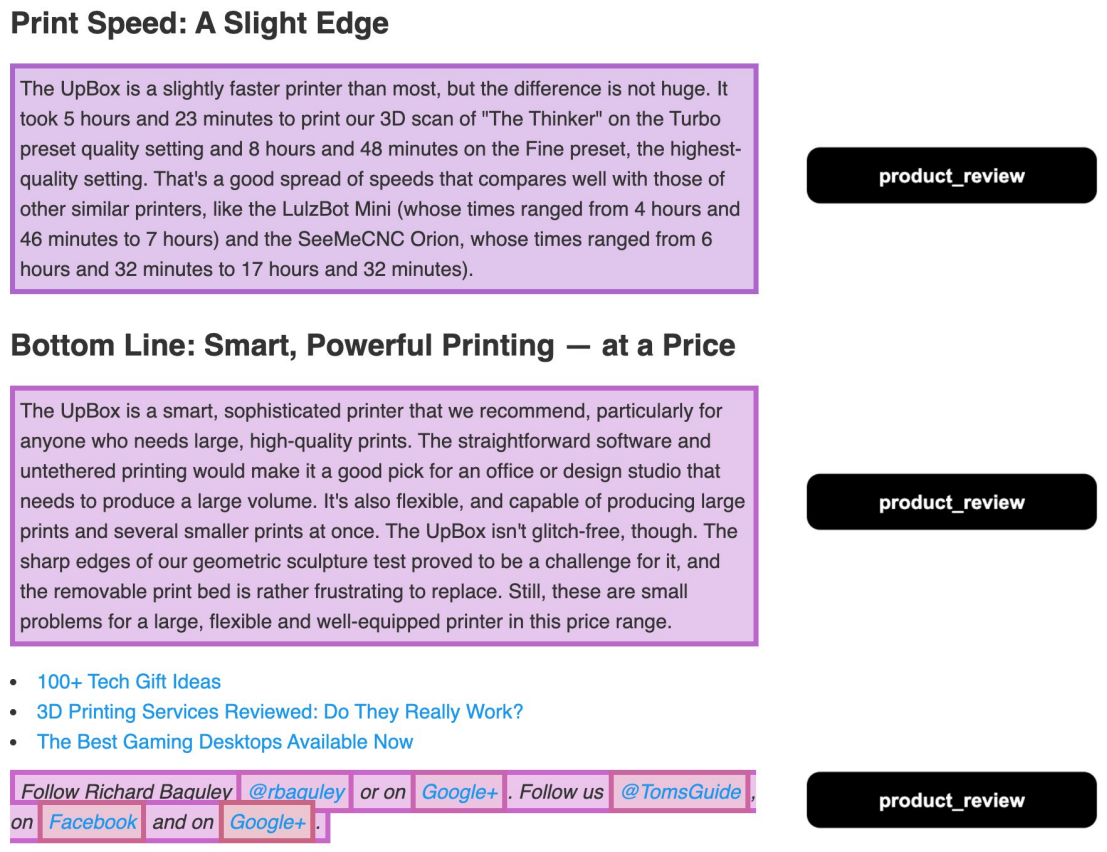

Figure 9: a bad case. The 3rd product review is mis-identified. The last paragraph should not be part of product reviews.